\documentclass[11pt]{amsart}
\usepackage[cp1251]{inputenc}
\usepackage[english]{babel}
\usepackage{amsmath}
\usepackage{amssymb}
\usepackage{amsfonts}
\numberwithin{equation}{section}

\begin{document}

\title[A note on a weakly singular elliptic equation]{A note on a weakly singular elliptic equation from theory of elasticity}
\author{Yu. A. Bogan}
\address{ Lavrent'ev Institute of Hydrodynamics, Lavrent'ev Prospect, 15, Novosibirsk, 630090, Russia}
\email{bogan@hydro.nsc.ru}

\begin{abstract}
Qualitative properties of a second order elliptic equation from the
 anisotropic elasticity are investigated. Some explicit
solutions for a disk are presented. Behaviour of these solutions in dependence
of coefficients is investigated. The problem of presence of  singularities of solutions at the origin of coordinates is discussed.
\end{abstract}
\maketitle

\section {Introduction}
To study the elastic behavior of cylindrically anisotropic materials began W. Voigt \cite{1} in the end of the 19th century. It was continued later by  Lekhnitskii \cite{2}, T.C. Ting \cite{3}, and J.Q. Tarn \cite{4}. W. Voigt was the first, who discovered, that in distinction from an isotropic material, some solutions for a cylindrically elastic material are singular at the origin of coordinates. In particular, boundedness of stresses at the origin depends on elastic constants.
Recall, that equations of the generalized Hooke's law are usually written in the orthogonal Cartesian coordinates; here they are given in the cylindrical coordinates.
This seemingly insignificant circumstance is nevertheless the source of difficulties when dealing with the cylindrical system of coordinates.
Moreover, all known solutions for a cylindrically anisotropic material are purely formal. Therefore, it is necessary to study the problem of solvability of principal boundary value problems for a cylindrically anisotropic material. This problem is studied below in the simplest set-up: anti-plane shear and the Dirichlet problem. It is shown below, that for constant shear modules this equation can be reduced to the Laplace's equation by a non-degenerate change of independent variables. Rewritten in the Cartesian coordinates, it is an elliptic equation, whose coefficients are only  bounded measurable functions.  A similar equation arises in stationary heat-conduction problems \cite{5}. Methods, developed in \cite{12}, can be applied to solve the basic elasticity problems for a cylindrically anisotropic solid.

\section{The anti-plane shear equation}

Recall (in short) derivation of the equation, describing anti-plane deformation (see details in \cite{2}). It is assumed, that there is a cylindrical rod of an arbitrary section, having the axis of cylindrical anisotropy parallel to its generator, that there is a plane of elastic symmetry, normal to the generator, and that elasticities in the cylindrical coordinate system $(r, \theta, z)$ do not depend on $z$. The origin of the cylindrical coordinate system is taken inside a body. Only two stress components (the tangential ones) are different from zero and independent of $z$. In these assumptions ($ u_r, u_\theta, w $ are displacements)

\[ \dfrac{\partial u_r}{\partial r}=0, \dfrac{\partial u_\theta}{\partial \theta} +u_r=0, \dfrac{\partial w}{\partial z}=0. \]
Then
\[ u_r= z( - \dfrac{1}{r}\dfrac{\partial w}{\partial \theta} + a_{45} \tau_{\theta z} + a_{55} \tau_{rz}) + c_1 \cos \theta +c_2 \sin \theta,\]

\[ u_\theta= z( - \dfrac{\partial w}{\partial r} + a_{44} \tau_{\theta z} + a_{45} \tau_{rz}) + c_2 \cos \theta -c_1 \sin \theta + c_0 r.\]
Here $c_0, c_1, c_2$  are real constants. Summands involving $ c_1,c_2 $ in previous formulas answer for a rigid displacement of a body; fix them somehow. If displacements do not depend on $ z $, the tangential stresses are expressed as
\begin{gather}
 - \dfrac{1}{r}\dfrac{\partial w}{\partial \theta} + a_{45} \tau_{\theta z} + a_{55} \tau_{rz} =0,\\ \nonumber
- \dfrac{\partial w}{\partial r} + a_{44} \tau_{\theta z} + a_{45} \tau_{rz} =0.
\end{gather}
Solve the previous equations with respect to stresses. Then
\begin{equation*}
\tau_{\theta z}= A_{44}\frac{\partial w}{r \partial \theta}+A_{45}\frac{\partial w}{\partial r},
\end{equation*}	
\begin{equation*}
\tau_{r z}=A_{45}\frac{\partial w}{r \partial \theta}+A_{55}\frac{\partial w}{\partial r},
\end{equation*}
where
\begin{equation*}
A_{44}=\frac{a_{55}}{\delta}, A_{45}=-\frac{a_{45}}{\delta}, A_{55}=\frac{a_{44}}{\delta}, \delta=a_{44}a_{55}-a_{45}^2 >0.
\end{equation*}
Here $ A_{ij}= A_{ij}(r,z), i,j=4,5 $; they have the physical meaning of elastic stiffnesses. Substitute this result into the unique equilibrium equation
\begin{equation*}
\frac{\partial \tau_{r z}}{\partial r}+ \frac{1}{r}\frac{\partial \tau_{\theta z}}{\partial \theta }+\frac{\tau_{r z}}{r}=f(r,\theta).
\end{equation*}
Then $w(r,\theta)$ is a solution of the equation
\begin{gather}
\frac{\partial}{\partial r}\biggl( A_{55}\frac{\partial w}{\partial r}+A_{45}\frac{1}{r}\frac{\partial w}{\partial \theta}\biggr )+\frac{1}{r}\frac{\partial}{\partial \theta}\biggl(A_{45}\frac{\partial w}{\partial r}+A_{44}\frac{1}{r}\frac{\partial w}{\partial \theta}\biggr) + \nonumber  \\
\frac{1}{r}\biggl( A_{55}\frac{\partial w}{\partial r}+A_{45}\frac{1}{r}\frac{\partial w}{\partial \theta}\biggr )=f(r, \theta).
\end{gather}
In particular, for a homogeneous rod with constant shear moduli and zero right-hand side, we get
\begin{equation}
A_{55}\biggl(\frac{\partial^2 w}{\partial r^2}+\frac{1}{r}\frac{\partial w}{\partial r}\biggr)+2A_{45}\frac{1}{r}\frac{\partial^2 w}{{\partial r}{\partial \theta}}+ A_{44}\frac{1}{r^2}\frac{\partial^2 w}{\partial \theta^2}=0.
\end{equation}

\section{The Dirichlet problem for a disk}
Consider as an example the Dirichlet problem for the equation (2.3)
in a disk of the radius $ R >0 $ with its center at the origin of a polar system. The boundary condition is
\begin{equation}
w(R,\theta)=g(\theta),
\end{equation}
where $ g(\theta) $ is a continuous summable periodic function of the polar angle. To make its solution unique require, as usual,  boundedness of $ w(r,\theta) $ at the origin.  Expand $ g(\theta) $ into the Fourier series with respect to $ \theta $, and put
\[ g(\theta)= \frac{a_0}{2}+ \sum_{n=1}^\infty( a_n\cos n\theta+b_n\sin n\theta), \]
\[ a_n=\dfrac{1}{\pi}\int\limits_0^{2\pi} g(\varphi) \cos n \varphi \,d\,\varphi, b_n= \dfrac{1}{\pi}\int\limits_0^{2\pi} g(\varphi) \sin n \varphi \,d\,\varphi, n=0,1. \ldots, \]
As usual, we apply the method of separation of variables. Then
\[ w(r,\theta)=f_0(r)+ \sum_{n=1}^\infty( f_n(r)\cos n\theta+g_n(r)\sin n\theta).\]
Functions  $ f_n(r), g_n(r) $ are solutions of the system
\[ A_{55}(\dfrac{d^2 f_n}{dr^2}+\dfrac{1}{r}\dfrac{df_n}{dr})-\dfrac{n^2A_{44}}{r^2}f_n+\dfrac{2n A_{45}}{r}\dfrac{dg_n}{dr}=0, \]
\[ -\dfrac{2n A_{45}}{r}\dfrac{df_n}{dr}+ A_{55}(\dfrac{d^2 g_n}{dr^2}+\dfrac{1}{r}\dfrac{dg_n}{dr})-\dfrac{n^2A_{44}}{r^2}g_n=0.   \]

Put $ f_n(r)=Br^{\lambda n}, g_n(r)=Cr^{\lambda n} $. Then $ B,C, $ and $ \lambda $ are solutions of equations
\begin{gather*}
 (A_{55}\lambda^2-A_{44})B+2A_{45}\lambda C=0,\\ -2A_{45}\lambda B+ (A_{55}\lambda^2-A_{44})C=0,
\end{gather*}
\[(A_{55}\lambda^2-2A_{44})^2+4A_{45}^2\lambda^2=0.  \]
The last equation can be decomposed as
\[ (A_{55}\lambda^2-A_{44}-2A_{45}i\lambda)(A_{55}\lambda^2-A_{44}+2A_{45}i\lambda)=0, \]
and it has the complex conjugate roots
$ \lambda_1=-i\alpha+\beta, \lambda_3=i\alpha+\beta, \lambda_2=-i\alpha-\beta, \lambda_4=i\alpha-\beta, $
where
\[ \alpha=-\dfrac{A_{45}}{A_{55}}, \quad \beta=\dfrac{ \sqrt{A_{44}A_{55}-A_{45}^2}}{A_{55}}. \]
Obviously, $ \beta>0 $, as it is assumed, that $ A_{44}A_{55}-A_{45}^2 >0 $. Roots $ \lambda_2, \lambda_4 $ are rejected as leading to the unbounded growth of a solution near the origin. If $ n=0 $, then $ f_0(r)=const $.
Now,
\[ f_n(r)=B_{n1}r^{(i\alpha+\beta)n} +B_{n3}r^{-(i\alpha+\beta)n},\]

\[g_n(r)=-iB_{n1}r^{(i\alpha+\beta)n} +iB_{n3}r^{-(i\alpha+\beta)n}.  \]
Having determined $ B_{n1}, B_{n3} $ from equations
\[B_{n1}R^{(i\alpha+\beta)n} + B_{n3}R^{(-i\alpha+\beta)n}=a_n,\]

\[-iB_{n1}R^{(i\alpha+\beta)n} +iB_{n3}R^{(-i\alpha+\beta)n}=b_n,   \]
we get, that
\begin{equation}
g_n(r)= (\dfrac{r}{R})^{\beta n}\{(a_n\sin(n\alpha \ln \dfrac{r}{R})+b_n\cos(n\alpha \ln \dfrac{r}{R})\},
\end{equation}
\begin{equation}
f_n(r)=(\dfrac{r}{R})^{\beta n} \{a_n\cos(n\alpha \ln \dfrac{r}{R})- b_n\sin(n\alpha \ln \dfrac{r}{R})\}.
\end{equation}
Then
\begin{gather}
f_n(r)\cos n\theta +g_n(r)\sin n\theta = \nonumber \\
\rho^n (a_n \cos n(\theta- \gamma \ln \rho) +  b_n \sin n(\theta- \gamma \ln \rho).
\end{gather}
Put $ \rho=(r/R)^\beta,  \gamma= \alpha/\beta $,
Then the formal solution of the Dirichlet problem is the series
\[ w(r, \theta)=f_0+ \sum_{n=1}^\infty (f_n(r)\cos n\theta+g_n(r)\sin n\theta), \]
which can be written as
\[ w(r, \theta)=f_0+ \sum_{n=1}^\infty \rho^n \{a_n \cos n(\theta- \gamma \ln \rho) +  b_n \sin n(\theta- \gamma \ln \rho)\}.    \]
It can be summed. Indeed, the sum
$ f_n(r)\cos n\theta +g_n(r)\sin n\theta$ is equal to
\[ \rho^n[a_n \cos n(\theta - \alpha \ln r)+ b_n \sin n (\theta -\alpha \ln r)].\]
Hence, the series
\[  u(\rho, \theta)= \dfrac{a_0}{2}+ \sum\limits_{n=1}^\infty \rho^n (a_n\cos (-n \gamma \ln \rho+ n\theta)+ b_n \sin (-n \gamma \ln \rho+ n\theta)) \]
for any bounded $a_n, b_n $ converges in the uniform topology in any internal point, as, since $ \rho <1$, it is estimated from above by the converging numerical series
\[M(1+\rho_1 + \rho_1^2 + \ldots +\rho_1^n+ \ldots), \]
where $M$ is the maximum of Fourier coefficients. Put
\[ u_n(\rho, \theta) = \rho^n (a_n\cos (n \alpha \ln \rho+ n\theta)+ b_n \sin (n \alpha \ln \rho+ n\theta)).  \]
Then
 \[ \dfrac{\partial u_n (\rho, \theta) }{\partial \theta^n}=\rho^n n^k (a_n\cos (n \alpha \ln \rho+ n\theta+ k \dfrac{\pi}{2})+ b_n \sin (n \alpha \ln \rho+ n\theta+ k \frac{\pi}{2})).  \]
Moreover,
\[\bigg| \dfrac{\partial u_n (\rho, \theta)}{\partial \theta^n}\bigg|\leq \rho^n n^k 2 M. \]
Any partial derivative with respect to $\theta$ of $ w(r,\theta) $ is estimated from above by a converging numerical series. Obviously, similar estimates are true for any partial derivative with respect to $ \rho $. The previous considerations in a natural way lead on to the introduction of new independent variables
\[z=\rho e^{i(\theta-\gamma\ln \rho)}, z=x_1+ ix_2, x_1= \rho \cos (\theta-\gamma \ln \rho), x_2= \rho \sin (\theta-\gamma \ln \rho).\]

Introduce also the expansion
\[f(z)= f(0)+ \sum_{n=1}^\infty c_n z^n.\]
\[ \quad c_n= \dfrac{a_n -i b_n}{R^{\beta n}}, n=1,2, \ldots. \]

Using previous expressions of the Fourier coefficients,  $u(\rho,\theta)$ can be written as the integral

\[ u(\rho, \theta)=\dfrac{1}{2 \pi}\int \limits_0^{2 \pi} g(\varphi)\{1+2 \sum_{n=1}^\infty \rho^n\cos n(\varphi-\alpha \ln \rho -\theta)\} \,d\,\varphi. \]
Put $\varphi +\theta -\alpha \ln \rho= \omega$. Then

\[1+2 \sum_{n=1}^\infty \rho^n \cos n \omega= \dfrac{1-\rho^2}{1+\rho^2 -2\rho \cos \omega} .\]

It follows, that  $u(\rho,\theta)$ is an analogue of the Poisson's integral for a harmonic function
\begin{gather}
u(\rho,\theta)= \frac{1}{2 \pi}\int \limits_0^{2 \pi} g(\varphi)\frac{1-\rho^2}{1-2\rho \cos(\varphi +\theta -\gamma \ln \rho)+\rho^2}\,d\,\varphi,
\end{gather}
where $ \gamma=\alpha/\beta $. In reality, it is a true Poisson integral for the harmonic function $u(\xi,\eta)$. It makes sense for any continuous function $ g(\theta) $. As $ \vert \rho \cos (\varphi-\theta -\alpha \ln \rho )\vert \leq \rho$, we are able to make the limiting transition $ \rho \to +0 $ and to derive the mean value theorem
\[ u(0, \theta)= w(0,\theta)=\dfrac{1}{2 \pi} \int\limits_0^{2 \pi}g(\theta)\,d \,\theta. \]
The fundamental solution of the equation
\[ \frac{\partial^2 w}{\partial r^2}+\frac{1}{r}
\frac{\partial w}{\partial r}+\frac{\lambda^2}{r^2}\frac{\partial^2
w}{\partial
\theta^2}=\frac{1}{r}\delta(r-r_0)\delta(\theta-\theta_0)\]

with pole at point $(r_0, \theta_0)$ has the form

\[w(r,\theta)=\frac{1}{2\pi \lambda} \ln\vert r^\lambda \exp i\theta
-r_0^\lambda \exp i\theta_0\vert.\]

The solution of the Dirichlet problem (3.1) has the finite Dirichlet integral $ D(u) $ (the bulk strain energy), if and only if

\[D(u)= \dfrac{\alpha^2+\beta^2}{\beta^2} \sum_{n=1}^\infty n(a_n^2+b_n^2)<\infty,\]
in other words, if boundary data $g(\theta) \in H^{1/2}(0,2\pi)$.
Here $ H^{1/2}(0,2\pi)$ is the fractional Sobolev space of the order $1/2$. It is well-known, that the Dirichlet problem for a disk has, in general, the divergent Dirichlet integral, if a boundary function is only continuous.

\section{On the existence of a weak solution of the Dirichlet problem}

Introduce the bilinear symmetric form
 \begin{gather}
a(w, \psi)=\int\limits_{\partial Q}\{(A_{45}\dfrac{1}{r}\dfrac{\partial w}{\partial \theta}+A_{55}\dfrac{\partial w}{\partial r})\dfrac{\partial \psi}{\partial r}+  \nonumber \\
(A_{44}\dfrac{1}{r}\dfrac{\partial w}{\partial \theta}+A_{45}\dfrac{\partial w}{\partial r})\dfrac{1}{r}\dfrac{\partial \psi}{\partial \theta}\} r\,d\,r \,d\,\theta
\end{gather}
and associated with it the quadratic $ a(w,w) $ one.
The equation (2.2) is the Euler equation of the quadratic form $ a(w,w) $.
Introduce the Cartesian coordinates $ (x_1, x_2)$, where $ x_1 =r \cos \theta, x_2= r \sin \theta $.Then the form $ a(w,w) $ shall be written as
\begin{equation}
a(w, w)=\int\limits_{\partial Q} [g_{11}(\dfrac{\partial w}{\partial x_1})^2+2 g_{12}\dfrac{\partial w}{\partial x_1}\dfrac{\partial w}{\partial x_2}+g_{22}(\dfrac{\partial w}{\partial x_2})^2] \,d\,x_1 \,d\,x_2,
\end{equation}
where
\begin{gather*}
g_{11}= A_{44}\sin^2 \theta -2A_{45}\cos \theta \sin \theta+A_{55}\cos^2 \theta, \\
g_{12}=(A_{55}-A_{44})\sin \theta \cos \theta+2 A_{45}(\cos^2 \theta -\sin^2 \theta), \\
g_{22}= A_{44}\cos^2\theta+2 A_{45}\cos \theta \sin \theta + A_{55}\sin^2 \theta.
\end{gather*}
The equation (2.2) will be written as
\begin{equation}
\dfrac{\partial }{\partial x_1}(g_{11}(x)  \dfrac{\partial w}{\partial x_1}+ g_{12}(x)\dfrac{\partial w}{\partial x_2})+ \dfrac{\partial }{\partial x_2}(g_{12}(x)  \dfrac{\partial w}{\partial x_1}+ g_{22}(x)\dfrac{\partial w}{\partial x_2})=f(x_1,x_2).
\end{equation}	
The coefficients of this equation are bounded measurable functions, discontinuous at the origin. Indeed, $ \sin \theta \cos \theta= x_1 x_2/(x_1^2 +x_2^2)$ is a well-known example of a discontinuous function. Assume, that the right-hand side $ f(x_1,x_2)\in L^2(Q) $, $Q$ is a bounded region on the plane $ (x_1,x_2) $ with a piece-wise smooth boundary, and there are constants  $C_1, C_2 > 0$, such, that
\[ C_1 |\xi|^2 \leq g_{ij}(x)\xi_i \xi_j \leq C_2 |\xi|^2 \]
for any $x \in Q$ and all real $ \xi=(\xi_1,\xi_2)$. A weak solution
of Dirichlet's problem is determined in the standard way as the function
$ w(x_1,x_2) \in H_0^1(Q)$, satisfying the integral identity
\begin{equation}
\int\limits_Q g_{ij} w_{x_i} \varphi_{x_j} \,d\,x=\int\limits_Q f(x)\varphi(x) \,d\,x
\end{equation}
for any  $ \varphi(x) \in C_0^\infty(Q) $. The existence of a unique weak solution of the Dirichlet problem for elliptic equations in the divergent form is well-known (see, for example, \cite{6}) in the assumption of uniform ellipticity and measurability of coefficients. Recall, that the previous equation means, that the  function $ w(x_1,x_2) $ assume (in some weak sense) zero value at the boundary. See, for example, \cite{6}, pages 237--241. It is obvious, that the previous equation belongs to this class, as the matrix $ g=(g_{ij}), i,j=1,2 $ is obtained by a rotation from the matrix $ A=(A_{ij}), i,j=4,5$ and has, therefore, the same eigenvalues and keeps up the property of positive definiteness.
It is worth to note, that similar elliptic equations arise in studying H\"{o}lder continuity of second order elliptic equations and serve as examples of equations with non-smooth solutions (see, for example, \cite{8}, \cite{9}, and \cite{7}).
Introduce new variables
\[ \xi= \rho\cos(\theta- \dfrac{\alpha}{\beta} \ln \rho ), \qquad \eta= \rho \sin(\theta - \dfrac{\alpha}{\beta} \ln \rho ), \quad \rho=r^\beta.\]

These are the characteristic variables simplifying the equation (2.2) in new variables and reducing it to the Poisson's equation
\[ \dfrac{\partial^2 u(\xi,\eta)}{\partial \xi^2}+\dfrac{\partial^2 u(\xi, \eta)}{\partial \eta^2}= f(\xi, \eta). \]
Here $u(\xi,\eta)=w(r, \theta)$. This family of characteristic curves has a singular point; indeed, the family of curves $\theta-\alpha \ln r = const$ is the family of logarithmic spirals with a singular point --- a spiral point at the origin of coordinates, according to the nomenclature of singular points of first order linear differential equations. If $\alpha =0$, the spiral singularity degenerates to a central one. The presence of a singular point of the field of characteristics leads, in general, to the singularity of a solution at the origin. Consider now the particular case $\beta=0$. The change of the
independent variable $\rho=r^{\lambda}$ reduces the homogeneous
equation (2.3) to the standard form of Laplacian, written
in polar coordinates: if we put $ w(r,\theta)=u(\rho,\theta) $, then
$ \Delta u(\rho, \theta)=0$, where $\Delta$ is the Laplace operator.
Hence $u(\rho,\theta)$ is a harmonic function with all nice
consequences, but, unfortunately, it is not so with respect to the
function $w(r,\theta)$. We consider the example
$w(r,\theta)=r^\lambda \cos \theta$. Then ${\partial w}/{\partial
r}=\lambda r^{\lambda-1}\cos \theta$. Hence, if $\lambda < 1$, it is
unbounded at the origin of coordinates. Of course, if $\lambda <2$,
second order derivatives are unbounded, when $\lambda <2$. Hence, if
$ \lambda \geq 2$, then second derivatives are bounded in any neighbourhood of the origin. Moreover, if $\lambda < 2$, second order derivatives of $w(r,\theta)$ are discontinuous at origin. Hence by the classical solution, when $\lambda <2$, one has to understand the function, satisfying the equation (2.3) inside a region with punctured origin. 
\section{Some comments about orthogonal coordinates}

First of all, recall the definition of orthogonal curvilinear coordinates. We will cite some properties from the English translation \cite{11} of the old Russian textbook on the differential equations. 

Suppose, that the point $ x $ is defined in terms of the three parameters $ \tau_j, j=1,3 $, that is, 
\[ x_k=x_k(\tau_1, \tau_2, \tau_3), k=1,2,3. \]
If these three functions, defining the coordinates of the point $ x $ in terms of the parameters $ \tau_i, i=1,2,3 $ are single-valued, then, to every set of values $ \tau_i, i=1,2,3 $ there corresponds one definite point $ x $. Suppose, that not only functions $ x_k(\tau_1,\tau_2,\tau_3), k=1,2,3 $ are single-valued, but  that they also have continuous partial derivatives. Examine the system of equations
\[ d x_i= \dfrac{\partial x_i}{\partial \tau_1} d \tau_1 +\dfrac{\partial x_i}{\partial \tau_2}d\tau_2 + \dfrac{\partial x_i}{\partial \tau_3}d\tau_3 \]
with respect to differentials $ d\tau_k, k=1,2,3 $. The determinant $ D $ of this system, which is composed of the partial derivatives $ {\partial x_i}/{\partial \tau_j}, i,j=1,2,3 $ is called the Jacobian of this system of functions; it is a function of the parameters $ \tau_i, i=1,2,3 $. The following proposition (the inverse function theorem) is known from the course of advanced calculus:

If the Jacobian of the system $ x_k=x_k(\tau_1,\tau_2, \tau_3), k=1,2,3 $ does not vanish in some neighborhood $ T $ of the values of the parameters $ \tau_i=\tau_i^0 $, to which the point $ x^0 $ corresponds, then, in some neighborhood $ X $ of the point $ x^0 $, the previous system admits a set of single-valued  inverse functions
\[ \tau_i= \tau_i(x_1,x_2,x_3), i=1,2,3,  \]
and these functions have continuous first derivatives with respect to $ x=(x_1,x_2,x_3) $ in the neighborhood $ X $ and they assume the values $ \tau^0=(\tau_1^0, \tau_2^0,\tau_3^0) $ at the point $ x^0 $. 

It is general practice to choose curvilinear coordinates to be in one-to-one correspondence with the points of the region to be studied, except possibly at certain points or lines where the Jacobian of the system of functions vanishes. These points (or lines) are called the singular points (or lines) of the corresponding coordinates.
Recall, that a coordinate system is called the orthogonal one, if the direction cosines of the tangent to the curve  $ \tau_j $ are proportional to the partial derivatives $ {\partial x_i }/{\partial \tau_j}, i=1,2,3 $, that is, the orthogonality condition
\[ \sum_{\alpha=1}^3 \dfrac{\partial x_\alpha}{\partial \tau_j}\dfrac{\partial x_\alpha}{\partial \tau_k}=0,  j \neq k.  \]

The displacement corresponding to the increment in the curvilinear coordinate $ d \tau_j $ is equal to
\[ ds_j=d \tau_j\sqrt{\sum_{i=1}^3 (\dfrac{\partial x_i}{\partial \tau_j})^2} = h_j d\tau_j, \]
where the quantities $ h_j, j=1,2,3 $ are called the Lame coordinate parameters. Thus, at the singular points of the coordinates, at least one of the L\'{a}me parameters vanish.

Consider now in more detail the cylindrical and the spherical coordinates. The cylindrical coordinates $ r,\varphi, z $ are defined by the system of equations
\[ x_1=r \cos \varphi, \quad x_2=r\sin \varphi, \quad x_3=z, (0 \leq \varphi \leq 2\pi). \]
The L\'{a}me parameters have the values $ h_1=h_3=1, h_2=r. $ Hence, on the cylindrical coordinate axis the parameter $ h_2=0 $, and, consequently, it is a singular line. Therefore, the coordinate $ \varphi $ does not have a definite value. Just the same can be told with the respect to the plane polar coordinate system. In other words, the  origin of the polar coordinate system is a singular point, as $ \varphi $ can take any value at the origin.

The spherical coordinates  $ r, \theta, $ and $ \varphi $ are defined by the system of equations 
\begin{gather*}
x_1 = r\sin\theta \cos \varphi, \quad x_2 = r\sin\theta \sin \varphi, \quad x_3 =r \sin\theta, \\
0 \leq \theta \leq \pi, \qquad 0 \leq \varphi \leq 2 \pi.
\end{gather*}
The L\'{a}me parameters have values
\[ h_1=h_r=1, \quad h_2=h_\theta =r, \quad h_3= h_\varphi= r \sin \theta. \]
The coordinate surfaces $ r=const $ form a system of spherical surfaces with a common center at the point $ x_1=x_2=x_3=0 $, called the spherical coordinate origin. The surfaces $ \theta=const $ form a system of circular cones with a common axis coinciding with axis 3 of the Cartesian coordinates. It is called the polar axis. Through every point not lying on the polar axis there passes one surface $ r=const $, $ \theta=const $, and one surface $ \varphi=const $. On the polar axis the parameter $ h_\phi=0 $; therefore, it is a singular line. On this axis, the coordinate $ \varphi $ does not have a definite value. At the singular point $ r=0 $, the coordinate $ \theta $ is also undefined. 

It is necessary to say some words on the treatment in \cite{1}, the Hooke's law in the cylindrical system of coordinates.
We shall use the Mir's translation \cite{1} of the original Russian edition of the book of Lekhnitskii. There are five pages --- 68-72, devoted to a discussion of properties of the generalized Hooke's law in cylindrical and spherical coordinate systems. We shall cite here one paragraph at the page 68, referring to the cylindrical coordinate system.

\textquotedblleft Here it is essential to make a very important remark. If the axis of anisotropy passes outside the body (for example, inside  a cavity), Eqs.(10.2) are not questionable. But if the axis of anisotropy passes through the body, there must necessarily be relations among different $ a_{ij} $ in a homogeneous body. Indeed, on the $ z $ axis coinciding with $ g $ there is no difference between the $ r $  and $ \theta $ directions, and all radial directions $ r $ must be equivalent not only among themselves, but also among all tangential directions $ \theta $ \textquotedblright.

Here equations (10.2) refer to the Hooke's governing equation, written in the cylindrical coordinate system, $ g $ is an axis of anisotropy, and $ a_{ij} $ are elastic coefficients. As result, he obtains eight equalities (10.4) between elastic coefficients. In particular, in antiplane shear at the axis of anisotropy $ g $ the equality $ a_{44}=a_{55}$ necessarily should be satisfied. The same remark is repeated on the page 72 with respect to the spherical coordinate system. He followed here the article of Voigt \cite{13}, pp. 507-508, where the assertion, given below, was written at first in the explicit way.

Compare the given above assertion with the discussion of general coordinates, given by Kellogg in \cite{4}. We cite the exercise 1 at the page 180:
\newline
\textquotedblleft Determine the points at which the functional determinant (3) is 0, in the case of spherical coordinates, and note that (a) at such points the coordinate surfaces cannot be said to meet at right angles, and (b) that such points do not uniquely determine the coordinates, even under the restriction of the usual inequalities $ 0 \leq \vartheta \leq \pi, 0 \leq \varphi \leq 2\pi $.\textquotedblright

It is obvious, that the cited assertion of Lekhnitskii and the exercise 1 of O. D. Kellogg are incompatible. O. D. Kellogg  was very skilled mathematician, a pupil of D. Hilbert, his volume on the potential theory is a classical reference, and we cannot believe, that this exercise is erroneous. Therefore, we agree with Kellogg, that directions are not determined at the origin, as the Jacobian is zero at the origin. The necessary properties of the cylindrical and spherical coordinate systems were recalled above. 

Consider, for example, the polar coordinate system.
The Jacobian $J$ of transition from  the orthogonal Cartesian system of coordinates $(x_1,x_2)$ to the polar one $(r, \theta)$ is
\begin{equation*}
J= \dfrac{\partial( r\cos \theta)}{\partial r}\dfrac{\partial (r\sin \theta)}{\partial \theta}-\dfrac{\partial (r\cos \theta)}{\partial \theta}\dfrac{\partial (r\sin \theta)}{\partial r}=r
\end{equation*}
and is zero at the origin, leading to the absence of one-to-one correspondence of points in the Cartesian and polar systems of coordinates; in other words, the mapping $ P=(r \cos \theta, r \sin \theta )$ is only injective in the whole plane without being surjective \cite{8}. See also the very interesting discussion of polar coordinates in \cite{8}; a modern version of the implicit function theorem is given there. Recall, that the latter, in particular, asserts that one-to-one correspondence of points is lost at any point where the functional determinant is equal to zero.
Probably, W. Voigt did not understand, that the origin has coordinates $(0,0)$ in the orthogonal Cartesian system of coordinates; and, meantime, in the polar system it has coordinates $(0, \theta)$, where $\theta$ is any number from the interval $(0,2 \pi]$.

Some more comments are  due. It is affirmed by some authors that very large stresses are not admissible in elastic problems. Of course,  boundedness of stresses is desirable in elastic problems, but it is not always possible to reach it even in seemingly simple situations. For example, modern composite- and nano--materials are, in general, strongly anisotropic. This circumstance necessarily leads to the appearance of boundary layer functions in an asymptotic expansion of a definite boundary value problem in powers of a small (or of great) parameter (see, for example, \cite{10}), and presence of large stresses in a small zone near a boundary of a region, for a given boundary value problem.

\end{document}